# Multimodal Alignment of Histopathological Images Using Cell Segmentation and Point Set Matching for Integrative Cancer Analysis

**Authors**:

Jun Jiang, Ph.D. 1
Raymond Moore, M.S. 1
Brenna Novotny, Ph.D. 1
Leo Liu, Ph.D. 1
Zachary Fogarty, Ph.D. 1
Ray Guo, M.D., Ph.D. 3
Markovic Svetomir M.D. Ph.D. 2
Chen Wang, Ph.D. 1
1. Department of Quantitative Health Sciences, Mayo Clinic, Rochester MN.
2. Department of Oncology, Mayo Clinic, Rochester MN.
3. Department of Laboratory Medicine and Pathology, Mayo Clinic, Jacksonville, FL.

**Abstract:**

Histopathological imaging is vital for cancer research and clinical practice, with multiplexed Immunofluorescence (MxIF) and Hematoxylin and Eosin (H&E) providing complementary insights. However, aligning different stains at the cell level remains a challenge due to modality differences. In this paper, we present a novel framework for multimodal image alignment using cell segmentation outcomes. By treating cells as point sets, we apply Coherent Point Drift (CPD) for initial alignment and refine it with Graph Matching (GM). Evaluated on ovarian cancer tissue microarrays (TMAs), our method achieves high alignment accuracy, enabling integration of cell-level features across modalities and generating virtual H&E images from MxIF data for enhanced clinical interpretation.

**Keywords:** Histopathology alignment, Histopathology registration, Bioimage analysis

**Introduction:**

As an important approach to reveal cell level details in cancer, histopathological images have been widely used in both clinic practice for diagnostic decision making and treatment follow up. Following different staining protocols, each modality of histopathology has its unique strength in highlighting specific aspects within tumor immune microenvironment (TIME). Among which, multiplexed Immunofluorescence (MxIF) images provide refined immune cell phenotyping, making it a favorable research tool for revealing cell behaviors in TIME. However, this imaging technique now is mainly used for research purposes due to the low reliability of marker signals caused by complex cyclic staining processes. On the other hand, H&E (Hematoxylin and Eosin) staining plays an irreplaceable role in providing standard clinical references by revealing cell morphology and texture patterns. Thus, accumulating studies have been combining cell level information from H&E with other histopathology stains to interrogate mechanism of cancer development and metastasis. For example, Bao etc. proposed a bifocal neural network to explicitly learn from H&E and IHC (Immunohistochemistry) for identifying abnormal in pathology images [1]. Gatenbee etc. [2, 3] compared the characterization of TIME using MxIF and H&E. Overall, the prerequisite of integrative multimodal histopathology TIME analysis is to establish cell scale correspondence between multimodalities, as the combined cell level characterization relies on co-localized cells within cancer tissues.

Although it is feasible to align diverse histopathological staining by annotating a few landmarks, this labor-intensive labeling work makes it hard to be applied to all the images in a large cohort study, especially for tissue micro-arrays (TMAs), as each whole slide image (WSI) contains hundreds of tissue cores. Image registration is an automatic technique that has been widely used for medical image alignment to establish pixel level correspondence. For example. Jonsson etc. [4] developed an image registration method targeted towards computer-aided voxel-wise analysis of whole-body PET-CT data. Jiang etc. developed an automatic alignment method by capturing the hierarchical nature of whole slide images [5]. Although image registration has been leveraged in histopathology images, it's still challenging to achieve cell-level alignment for the following reasons. First, the size of histopathology image (usually more than 3GB) is much larger than other medical images, such as CT or MR (usually less than 1MB), which means many existing methods can't be directly applied in our scenario due to the limitation of computational resource; Second, the contents (cells) within histopathology imaging are dense and small to the image size, which exhausts the algorithm to converge; Third, the content of images acquired from diverse modalities could be dramatically different. Some cells in one modality can't find the compartments in another modality even if the images are from the restained tissue section, as the pigment only bond to cells with specific marker. In consequence, algorithms may not be able to find enough reliable landmarks to establish the location correspondence. Thus, there exists an urgent need for multimodal histopathology alignment to enable integrating detailed information from different stains.

Given the challenges of image alignment by directly using pixel information, it's straightforward to consider using cell segmentation outcomes from histopathology images as the starting point to develop approaches for establishing cell spatial correspondence. This resolving path does not cause extra cost to the existing workflow since cell segmentation is the premise of cell level downstream spatial analysis. On the contrary, it can significantly reduce computational complexity as the number of cells is dramatically less than the number of pixels within a histopathology image. Along this resolving path, two major components are required: 1) cell segmentation; 2) point set alignment. Cell segmentation denotes the methods detect cells and locate the boundaries of cells from histopathological images. With the advancement of AI and Deep Learning, many cell segmentation models achieved not only superior performance but also high generalizability and robustness[6-8]. Since cell segmentation is comparable to the key points detection steps within the traditional image registration process, the advancement in AI-enhanced cell segmentation can be fully leveraged in this solving path. With the centroids of cell segmentations, cells within histopathology image can be summarized as a point set. Thus, establishing cell spatial correspondence can be formulated as point sets alignment problem, which has also been widely explored within machine learning and pattern recognition field [9, 10].

In this paper, presented a novel framework to establish spatial correspondence for cells from different histopathology stainings, allowing us to validate cell features across different modalities and enabling the synergy of combinational analysis. By formulating the cell spatial correspondence into point sets alignment, we used cell segmentation results from histopathology image pairs as the basis, cells were first aligned with Coherent Point Drift (CPD) [10] and then the alignment was calibrated using Graph Matching (GM)[11]. We systematically evaluated the alignment accuracy for both restained and serial section images from ovarian cancer TMA images. The proposed method achieves encouraging alignment accuracy and portable to different cell segmentation methods. Within the aligned multimodal image pairs, we found that cell morphology and nucleus staining features are highly correlated, especially in restained tissue cores. Regional analysis also confirmed high concordance in cell density and cell composition despite discrepancies in automatic cell segmentation results within different modalities. Moreover, the multimodality alignment enabled training a deep learning model which generates high fidelity virtual H&E, providing clinical reference for MxIF images.

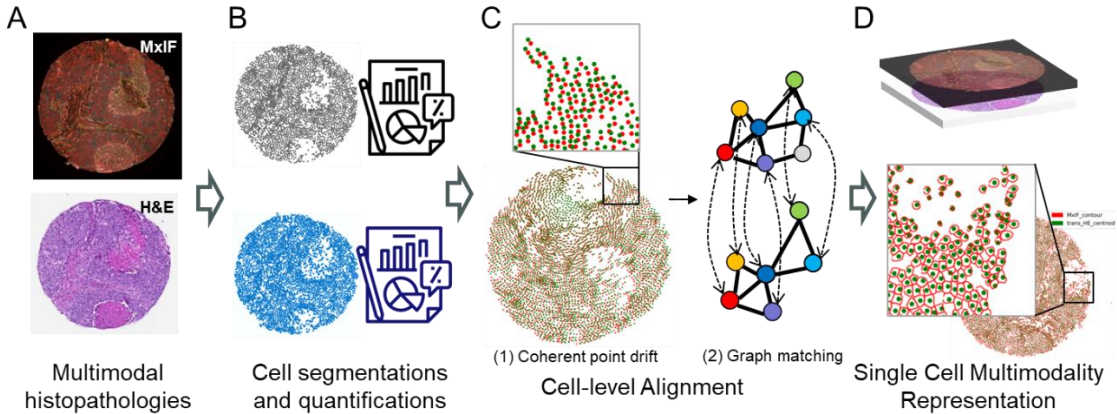

Figure 1. Overview of our framework. A) MxIF and H&E images to be aligned; B) Cell segmentations and quantifications; C) Cell-level alignment. 1) coherent point drift. 2) graph matching; D) Aligned MxIF and H&E images with combined cell-level representations.

**Methods:**

1. **Dataset and materials**
**Slide preparation, image acquisition and preprocessing.**

The dataset used for this study consists of three ovarian tissue micro array (TMA) images from two tissue sections (Supplementary Figure 1). To create this TMA, ovarian tissues were retrieved from Mayo Clinic Ovarian Tissue Archive and punched into arrays and made as formalin-fixed paraffin-embedded (FFPE) blocks. Two continuous 5μm sections were cut for MxIF and H&E staining.

For the first section (Section 1 in Supplementary Figure 1), a sophisticated approach involving cyclic MxIF (Multiplexed Immunofluorescence) staining was employed to meticulously assess protein expression levels within individual cells. This method enabled iterative quantification, wherein tissues were sequentially subjected to fluorescent antibody staining, digitally scanned, and then the dye was inactivated to facilitate successive rounds of staining and imaging. The raw MxIF images from CellDive were processed using our previous preprocessing pipeline to remove auto-fluorescence (AF) and convert into OME tiff file. Following the completion of MxIF imaging, the restained section was restained with H&E (Hematoxylin and Eosin), providing additional morphological context to complement the molecular insights garnered through MxIF. The second section (Section 2 in Supplementary Figure 1) only underwent H&E staining. Specifically, the MxIF images were obtained from CellDive, and H&E images were from Zeiss Axio Z1. Images from both sections were used in our method development and evaluations.

Since the TMA images consists of many FOVs (field of views) and each FOV is a tissue core image, the TMA images were de-arrayed within QuPath [12] for downstream alignment and evaluation. An ID was assigned to each tissue core. The extracted TMA core images were visually inspected and the tissue cores without sufficient tissues were discarded. StarDist was used as our baseline for cell segmentation. The segmentation results were imported into QuPath for quantification. To investigate how the cell segmentation results affect the alignment, we also include watershed threshold segmentation method for comparison.

2. **Alignment Algorithms**
A. **Problem Formulation**

Considering the scenarios in histopathology image scanning, three transformations were included in for alignment, in which scaling results from the differences of pixel spacing within two images, rotation and translation result from the differences to the tissue locations. The overall transformation from one image to another can be summarized into a transformation matrix $M$.

$$M = \begin{bmatrix} S*\cos(\theta) & -\sin(\theta) & dx \\ \sin(\theta) & S*\cos(\theta) & dy \\ 0 & 0 & 1 \end{bmatrix} \quad (1)$$

In this formular, the scaling was denoted to $S$, rotation was denoted to $\theta$. The translation/shifting distances on $x$ and $y$ directions were denoted to $dx$ and $dy$, respectively. The distance of translation $T$ can be calculated as $\sqrt{dx^2 + dy^2}$.

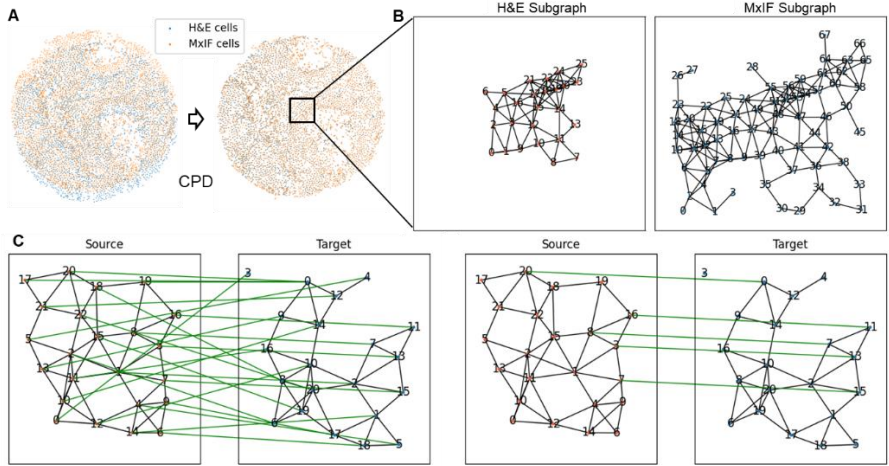

Figure 2. Illustration of alignment algorithms. (A) Cell centroids were first aligned with CPD, and then fine-tuned with GM. Within GM phase, (B) subgraphs were created for H&E and MxIF subregions with raw aligned cells as graph nodes. (C) The putative matching nodes were then filtered to establish the correct cell correspondence.

### B. Coherent Point Drift (CPD)

CPD was employed to generate a raw alignment for cell centroids (Figure 2A). This method formulates the point-set registration as a probability density estimation problem, where one point cloud is represented using a Gaussian Mixture Model (GMM). Considering the transformation defined in formula (1), the original CPD algorithm needs to update three major parameters ($\theta, S\ and\ T$) for the alignment. As an Estimation-Maximization (EM) method, CPD iteratively estimates the point correspondence and updates the three parameters until convergence [10].

Specifically, to reduce the degree of freedom within the CPD algorithm, the cell centroid coordinates were converted into micrometers (μm), so the scale value within formular (1) can be set to constant value ($S = 1$). In our case, the pixel size of MxIF and H&E image are 0.325μm and 0.212μm, respectively. Pixel coordinates can be converted to micron units by multiplying pixel size.

### C. Graph Matching (GM)

Since the CPD-based method estimates the point correspondence using GMM, the alignment was estimated based on the cell densities rather than one-to-one correspondence. To calibrate the alignment, we formulated the cell-to-cell correspondence to a graph matching problem, which was implemented as a modular concatenated to the CPD in our framework (Figure 1C). In this phase, the cells from H&E or MxIF were abstracted as nodes in a mathematical graph.

**C1. Creating subgraphs**

After the CPD transformation, the cells from H&E or MxIF were roughly aligned. Two subgraphs, the source graph from H&E and the target graph from MxIF were created for matching. Since the subregions with denser cell counts usually presents magnificent tissue architectures, while the regions with sparse cells can more likely be affected by cell segmentation noise [13], the cells for graph creation were randomly sampled from subregions with dense cell distributions. Kernel density estimation (KDE) was used to calculate density value d for each cell, cells with d>=0.5 were optional cells for sampling. To create a source graph for H&E, the cells were sampled from a window with size of 50 µm (Figure 2B). For the target graph in MxIF, the location of the sampling window was calculated by mapping the centroid of sampling window in H&E using CPD transformation, and the window size was set to 150 µm. The edges for both source and target graphs were added based on proximity of cells if the pairwise distance was less than 15 µm.

### C2. Node matching

Graph matching is a computational technique that seeks to establish node-to-node correspondence between two or more graphs. This process involves solving a combinatorial optimization problem, which is typically NP-hard. The goal is to match nodes from one graph to another while considering the similarity of the nodes (node affinity) and the relationships between connected nodes (edge affinity). By considering both types of affinities, graph matching methods tend to be more resilient to noise and outliers in the data, making them suitable for complex real-world applications.

In our case, the affinity matrices were built by calculating node and edge similarities between corresponding graphs. Node features, such as cell morphologies (perimeter, solidity), were used to quantify node affinity. Edge affinity was based on the distances between connected nodes, with edge features capturing the structural relationship between node pairs in the graph. Reweighted Random Walks Matching (RRWM) solver was then used to iteratively refine the matching score based on the affinity matrix, which yields a probability distribution for potential matches. This soft matching results were converted into a final "hard" matching using a combinatorial optimization method like the Hungarian algorithm or the Sinkhorn algorithm[11, 14].

### C3. Filtering matched node pairs

Although graph matching establishes the one-to-one correspondence between nodes in source and target graphs, not all the node correspondences are true positive. Considering the cells segmented out of the original H&E and MxIF images also representing the microstructures of ovarian tissue images, we introduced Locality Preserving Matching (LPM)[15] to exclude unreliable matching node pairs for the final spatial correspondence, as this method was explicitly designed to filter putative matching point pairs while preserve the spatial neighborhood relationship among points. Figure 2C illustrates the graph matching results before and after LPM filtering, only the matching pairs preserves the neighborhood relationships were kept. In our case, the graph nodes in source plot were the cell centroids from H&E segmentation, while the nodes in target plot were cell centroids from MxIF. After the matching point pairs filtering, the left point pairs were used to calculate the affine transformation that translocate H&E cell locations to MxIF cell locations. To implement our framework, three major python packages StarDist [16], probreg [17] and pygmtools [11] were used. More details about the implementations can be found in our shared GitHub code (https://github.com/dimi-lab/MultimodalityHistoComb).

3. **Alignment Accuracy Evaluation**
A. **Creating alignment ground truth**

The ground truth of cell level correspondence was established by annotating landmarks within H&E and MxIF image pairs from 20 TMA cores without significant tissue damage. For each selected pair of tissue core, MxIF and H&E images were imported into QuPath, in which the point annotation tool was used to annotate eight pairs of landmarks (Supplementary Figure 2). To efficiently locate corresponding cell pairs within H&E and MxIF, significant microarchitectures, such as interface of tumor and stroma, were visually referred to navigate to individual cells. Landmark points were placed at the center of the cells close to significant microarchitectures in both H&E and MxIF. To minimize annotation errors, lymphocytes were preferred as landmarks, since their nucleus are dark and small. Then the coordinates of annotated landmarks were exported from QuPath to calculate the ground truth transformation matrix M, in which the rotation $\theta$, scale $S$ and translation $dx$ and $dy$ for the rigid alignment were summarized in formular (1).

### B. Evaluation metrics

To quantitatively evaluate the performance of alignment accuracy, three metrics were used by referring to existing work [2], as summarized in formular (2). We compared the differences between alignment results from annotation and our methods. Following the completion of alignment, the location of landmarks annotated in the source image were mapped to the new locations within the target image, labeled with red dots in Figure 3 A. The average distance between landmarks in target image (green dots) and transformed landmarks from source image (red dots) was used to measure the overall alignment accuracy, denoted to $\Delta D$ in (2). With the annotated landmarks, the ground truth rotation angle $\theta$ and translation distance $T$ can be calculated. The difference between the results from our method and the ground truth were the other two quantitative metrics, denoted to $\Delta\theta$ and $\Delta T$ respectively. Using the defined evaluation metrics, alignment accuracies were evaluated for re-stained and serial sections.

$$\Delta D = \frac{1}{n}\sum_{i=0}^{n}|d_i - d'_i|$$
$$\Delta T = |T' - T|$$
$$\Delta\theta = |\theta' - \theta|$$
(2)

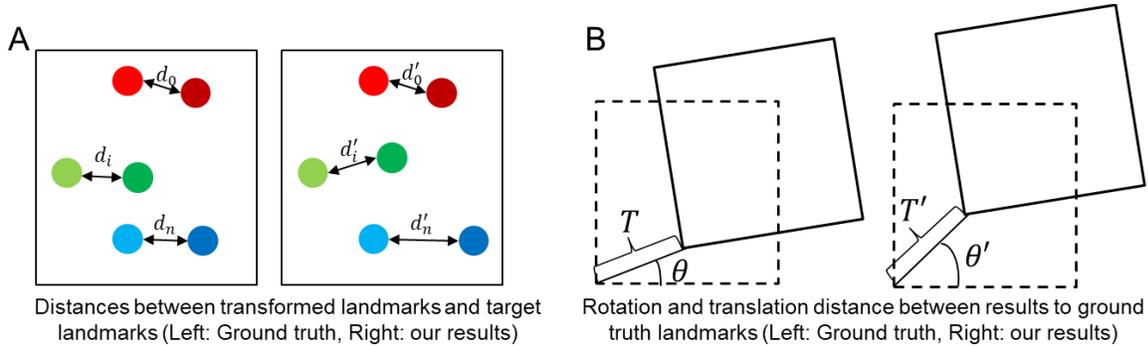

Figure 3. Illustration of the evaluation metrics. A) Distances between landmarks after transformation (red dots) and target landmarks (green dots). B) Rotation and translation. Landmark distances, translations and rotations were denoted to $d_i, T \ and \ \theta$ in ground truth and $d'_i, \theta' \ and \ T'$ within our results.

### 4. Downstream Analysis

Once the images were aligned, many downstream analyses were enabled by providing complimentary information or validation. With our ovarian TMA dataset, since H&E usually serves as clinical diagnosis reference, we demonstrated that our alignment helps to 1) validate cell features from MxIF and 2) facilitate training a deep learning model that generate high fidelity of virtual H&Es using MxIF image channels.

### A. Cell feature concordance

Pathologists make diagnoses based on tissue staining and cell morphology, thus automatic cell feature extraction process is essential to simulate pathologists' interpretable diagnosis for large cohort study. By examining colocalized cell nucleus staining and morphology features, we can assess cell and regional levels feature consistency, ensuring the quality of MxIF data and reliability of subsequent analysis. To this end, cells were identified with segmentation models [7] and quantitative cellular features were extracted within QuPath [18] for MxIF and H&E respectively. Measurements like the average DAPI/Hematoxylin signal intensity per cell and the area of a cell nucleus were compared by cell or region for restained and serial sections.

B.  Generating virtual H&E from MxIF

To review the dark field MxIF images, H&E counterpart is preferred to provide cell level references. In practice, it is tedious work as serial section from the same tissue block needs to be cut and stained. Moreover, within the obtained images, cells are not one-to-one corresponded, leading to unreliable clinical reference. Compared to serial sections, the most ideal situation is restain the same tissue with H&E after MxIF scanning. same or serial sections). Although many MxIF platforms provide virtual H&E, the quality is unacceptable for clinical interpretation. Using our alignment method and dataset, we presented the feasibility of generating high fidelity virtual H&E using some image channels from existing MxIF by training a deep generative model. Without extra tissue cut, staining and scanning process, the generated images were visually inspected by pathologists, and downstream cell detection was conducted to assess the quality of generated virtual H&E.

**Results:**

1.  **Our framework achieves cell scale alignment for MxIF and H&E images.**

Our method was tested on both restained and serial sections. For each section and each core, the H&E images were aligned to MxIF. According to the landmark annotations, ground truth transformation *M* was calculated and applied to the restained and serial section. The ground truth results H&E were visualized in Figure 4A column 2 and 4, while the results from our method were visualized in column 3 and 5. For both sections, our H&E results were visually aligned well to the corresponding ground truth H&E, even for the breaking up tissue core, shown in the first row of Figure 4A. Meanwhile, our H&E results for both sections aligned well to the MxIF, even though there were micro architecture and staining color differences between the two sections.

Taking a tissue core as an example (core ID: B-11), to visually check the alignment at cell level, the landmarks from H&E (red dots) were translocated and plotted together with landmarks from MxIF (blue dots), as showing inf Figure 4 B and D for section 1 and section 2 respectively. For the restained section (Figure 4 B), the MxIF landmarks were fully overlayed on translocated H&E landmarks, indicating minor alignment error. The automatic alignment errors were higher than the ground truth alignment, especially for serial section. According to the quantitative evaluation metrics, the landmark distance differences $\Delta D$ in Figure 4C are comparable ($\Delta D_{sec1} = 32.47 \pm 27.25 \ vs. \ \Delta D_{sec2} = 34.47 \pm 25.71$) ) to that in Figure 4E, suggesting that with manual annotation, both restained and serial sections can be aligned well. Compared to the average values for restained section ($\Delta T = 6.94 \pm 4.73$ and $\Delta \theta = 1.26 \pm 0.98$), serial sections showed higher values ($\Delta T = 20.77 \pm 17.46$ and $\Delta \theta = 2.15 \pm 0.97$), indicating the rotation and translation errors were higher.

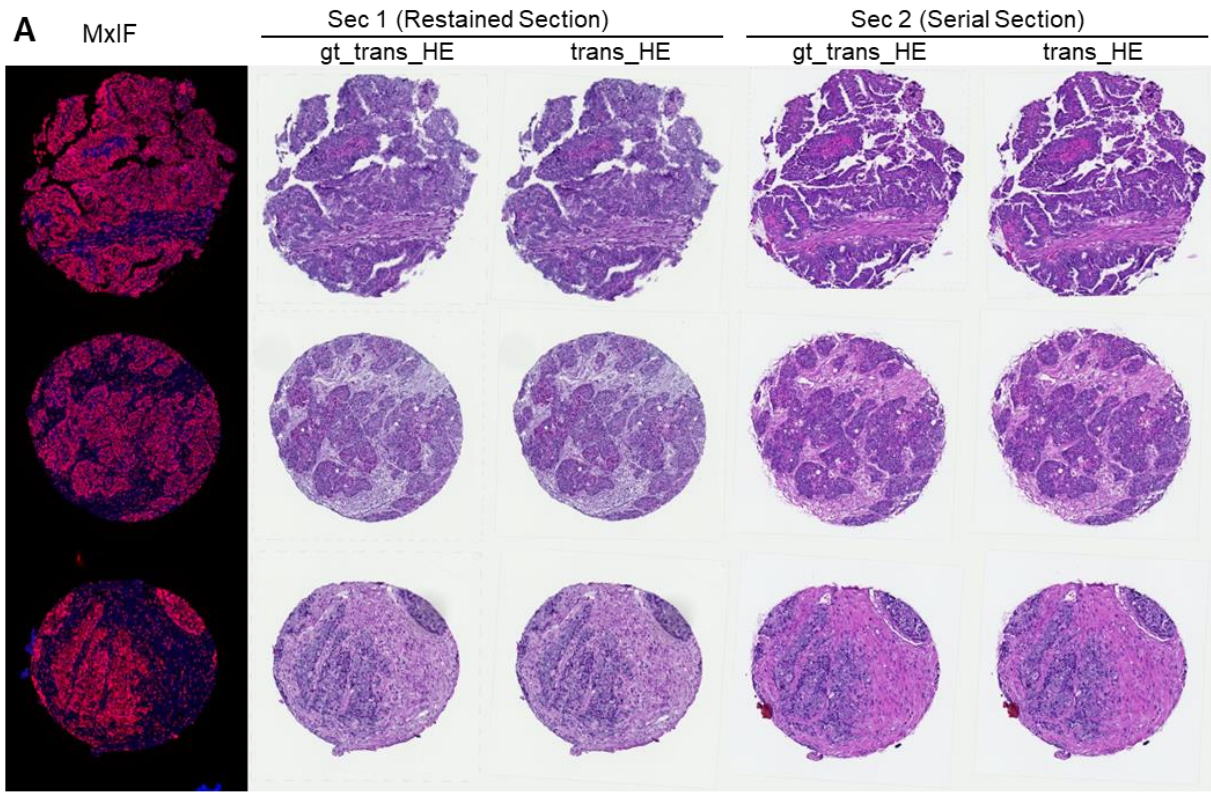

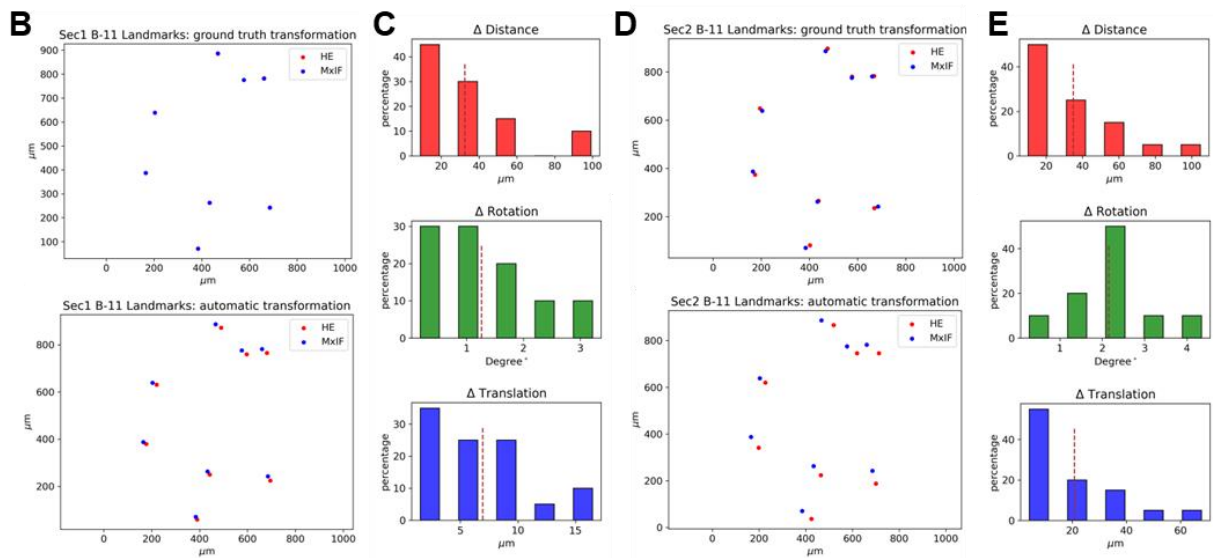

Figure 4. Qualitative and quantitative evaluation results. A) Examples of MxIF and aligned H&E for both restained and serial tissue sections. B) A restained section example of aligned landmarks obtained from annotation (ground truth) and our (automatic) method. C) Quantitative evaluation for restained section. Histogram for evaluation metrics in formular (2) (∆D, ∆θ and ∆T from top to bottom), with dash line labeling the mean value. D) A serial section example of aligned landmarks obtained from annotation (ground truth) and our (automatic) method. E) Quantitative evaluation for serial section. Histogram for evaluation metrics in formular (2) (∆D, ∆θ and ∆T from top to bottom).

**2. Our framework is portable to different segmentation models.**

Since our method relies on cell segmentation, we investigated the alignment performances with respect to two different segmentation methods, StarDist and Watershed. As the restained tissue section provide more reliable one-to-one cell correspondence, our evaluations were conducted on Section 1. Although Watershed generates more over- and under- segmentation cell instances [19], similar transformed H&E image can be obtained by applying the transformation from our method. Taking a same tissue core (ID: B-11) as an example, the discrepancies between aligned H&E results images (Figure 4A and 5A) can only be aware of by zooming into cell level by visualizing the cell centroids before and after transformation (Figure 5B). Moreover, according to the quantitative evaluation metric ∆D, the performance of using Watershed was even better than that using StarDist (Figure 5C).

We also tested the method using hybrid results from two cell segmentation models, with StarDist for MxIF and Watershed for H&E. Compared to the default segmentation method StarDist, evaluation metric $\Delta D$ was lower when using hybrid cell segmentation results for alignment with $\Delta D_{StarDist} = 32.47 \pm 27.25 \; vs. \; \Delta D_{hybrid} = 15.87219264 \pm 14.65349468$. The performance fluctuations showing in Figure 4C,E and Figure 5C,D can be result from the inconsistencies within H&E and MxIF cell segmentation. Since there were many cell segmentation models available specifically for H&E or MxIF, our method provided a straightforward and portable way to align cells using segmentation results as the starting point for alignment.

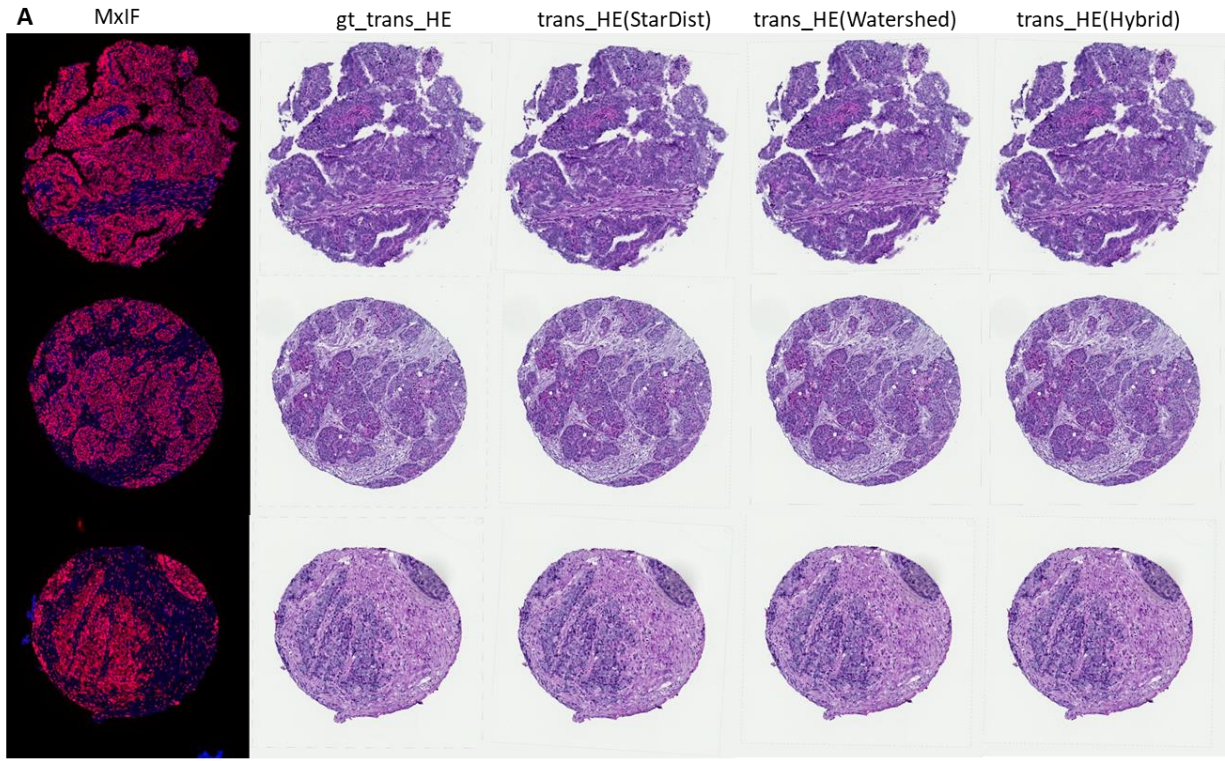

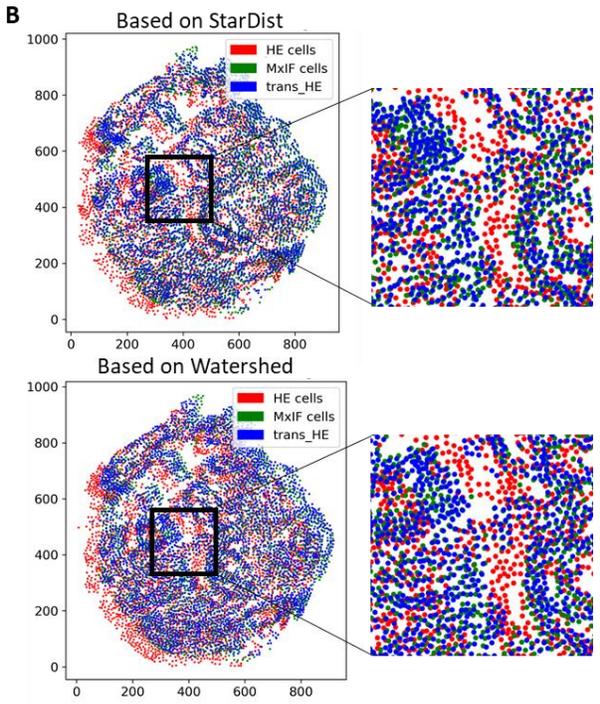
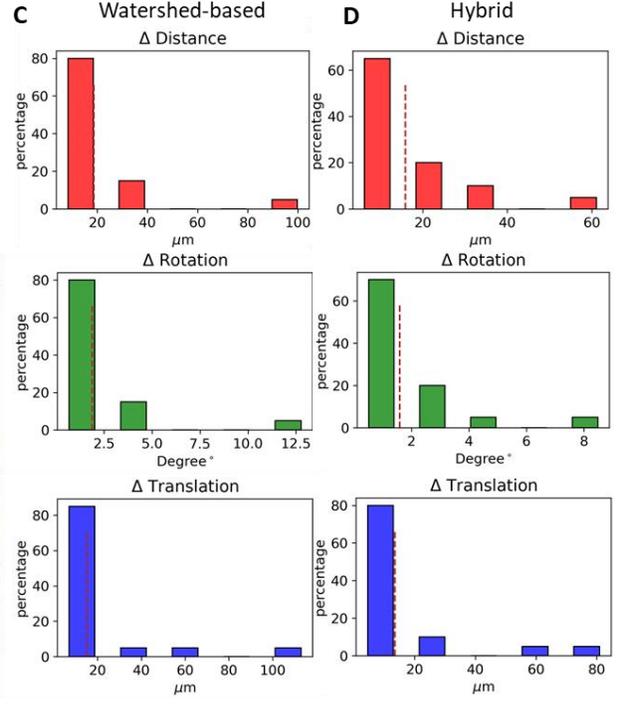

Figure 5. Results of alignment initiated from different cell segmentation methods. A) The alignment results for restained section (Sec1). B) Cell centroids scatter plot for MxIF, original H&E and transformed H&E. C) Quantitative metrics for alignment accuracy, using Watershed segmentation results as the starting point. D) The same quantitative metrics. For MxIF, used the StarDist results as the starting point; For H&E, used the Watershed results as the starting point.

## 3. Multimodality alignments enable cell and regional feature concordance evaluation.

Since our alignment framework aims to enable integrative spatial analysis within the aligned tissue space, cell and regional signal concordances after the alignment were evaluated. By applying the transformation to the cell segmentation results, cells from H&E and MxIF can be aligned to the same space, as shown in Figure 5B. However, even if we apply the ground truth transformation, rigid one-to-one cell correspondence cannot always be established for two reasons: 1) The cell within tissue section images were not the identical, especially for serial sections. Ideally, for the restained tissue, the cells should be the same for different imaging techniques. However, there could be cell being washed off or even large piece of tissue damage during the stain-restain process. 2) The cell segmentation models don't always generate the same segmentation results from different modalities as the models were trained from and applied to different modalities independently. For example, there are pretrained StarDist models available for H&E and MxIF respectively, but the models were trained from different dataset. The results will be different when we apply the model to H&E and MxIF independently.

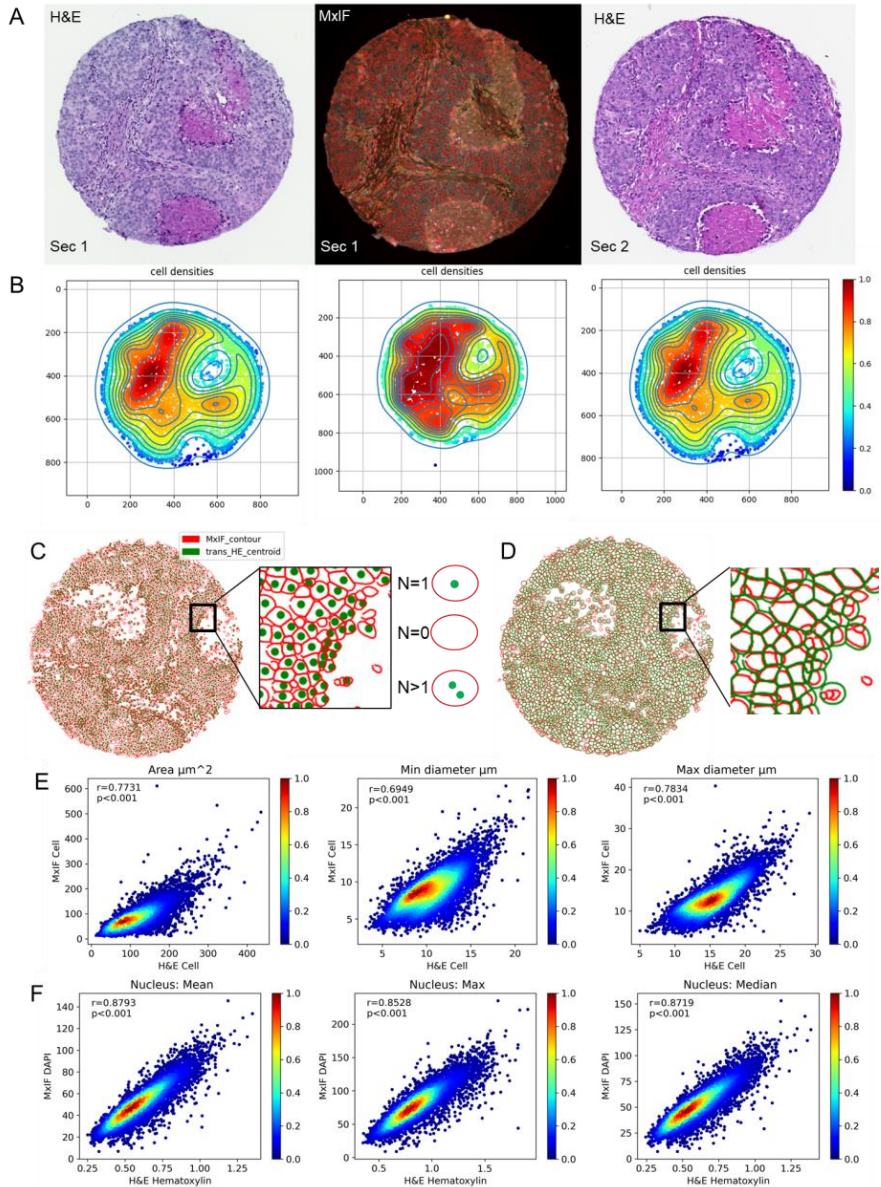

Figure 6. Visualization of aligned cells and cell feature concordance evaluations. A) An example of aligned cells with cells from MxIF (red) showing cell contours while the cells from H&E(green) showing cell

centroids. B) The same example with cells from both H&E (green) and MxIF(red) showing cell contours. C) Concordance of morphological features: Cell Area, minimum diameter, and maximum diameter. D) Concordance of Hematoxylin (H&E) vs. DAPI (MxIF) staining features in nucleus were evaluated with mean, median and maximum.

To check the single cell level feature concordance between H&E and MxIF, the restained tissue section was employed. By visualizing the aligned cells into the same space, as shown in Figure 6A,B, cells from two modalities demonstrated three scenarios in single cell level correspondences: 1) N=1, the cells are one-to-one corresponded; 2) N=0, a cell detected in one image didn't corresponds to any cell in another image; 3) N>1, there could be more than one cell in one image corresponding to a cell in another. According to our observations, for both N=0 and N>1, the mismatch was caused by segmentation inconsistency between H&E and MxIF. The StarDist model we applied to MxIF was sensitive to DAPI signals, which led to two consequences: 1) For cells with weak DAPI signals, the instances detected in MxIF could be recognized as fragments within H&E which can be filtered out within segmentation process (N=0). 2) For cells close to each other and with strong DAPI signals, multiple instances that recognizable in H&E can be detected as single cell in MxIF (N>1) (supplementary figure 3). Since N=0 and N>1 were extreme cases for cell segmentation, we investigated the morphological and staining feature concordance with one-to-one cell correspondence (N=1) within all the 20 evaluation cores. As shown in Figure 6C, morphological features, including cell area, minimum diameter, and maximum diameter, from H&E cells and MxIF cells were highly correlated. Meanwhile, the staining concordances were compared on DAPI and hematoxylin as both are cell nuclei pigments (Figure 6D). For both morphological and staining features, r and p values for Pearson correlation were calculated and shown in the plots.

To be more generic, we investigated the cell level feature concordance under two different conditions: 1) align with manually labeled landmarks vs. our automatic alignment; 2) serial vs. restained section. Each tissue core was also aligned to randomly chose core as the baseline for comparison. Specifically, after alignment, each cell in H&E was corresponded to the nearest MxIF cell for feature comparison. As shown in Figure 7A and 7B, both morphological (cell area) and staining features (DAPI vs. Hematoxylin) demonstrated higher feature concordance than the serial section. Though the concordance based on our alignment was lower than the ground truth, it's comparable in both restained and serial section, and were significantly higher than random alignment.

Although the cell-level feature concordance in serial sections was lower, similar tissue structures in H&E and MxIF were observed, as showing in Figure 7C,D,F,G. Cells within H&E and MxIF were classified to tumor vs. non-tumor by referring the cell features extracted by StarDist models and QuPath[19, 20]. Then, the concordance of cellular composition was evaluated by comparing proportions of tumor cells region by region. As shown in Figure 7C, heatmap was used to demonstrate the reginal similarity of cell composition, in which warm color indicates high concordance. The most disaccord region

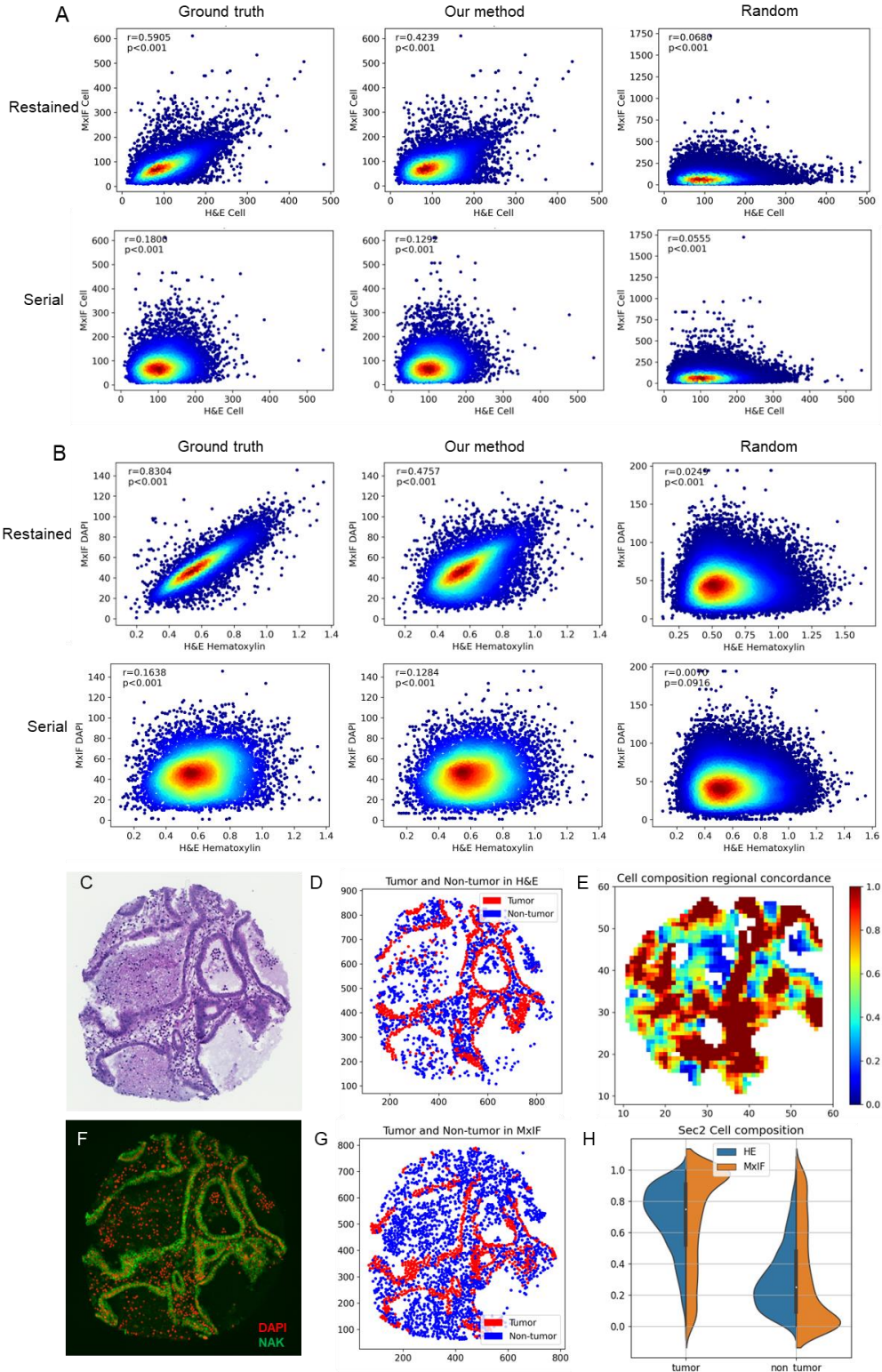

Figure 7. Feature concordance and cellular composition comparison. A) Cell area concordance under different conditions. B) Cell nucleus staining concordance under different conditions. Row 1: restained section; Row 2: serail sections; Column 1: aligned with manual annotation; Column 2: aligned with our method; Column 3: randomly align to a tissue core. H&E (C) and MxIF (F) images for an example TMA core, and the automatic cell classification results (D, G). E) regional cell composition similarity of the example. Lower value within heatmap means low concordance. High value indicates high concordance. H) The distribution of cell composition for all our serial tissue cores.

**Discussions**

We presented a novel histopathology image registration framework based only on cell segmentation results, which are the prerequisites for many downstream analyses. This setting enables the possibility of combing cell features from multiple histopathology modalities without introducing extra burden for modifying the existing analysis pipeline for each modality. Developed on a specific dataset, we demonstrated not only the encouraging alignment accuracy, but also the pitfalls that may occur under different tissue cut and staining settings.

Although there are existing approaches consider the histopathology image alignment as a non-rigid registration work because of tissue warping [21], we simplified the TMA image alignment as rigid image transformation. Because, first, comparing to the whole tissue slide image, the size of each TMA core image is small (~40,000 pixel in height vs. ~4000 in height). According to our observation, tissue warping can be more significant on large pieces of tissues, but not on TMA cores. Second, significant tissue warping is a quality issue that should be avoided in slide preparation stage for reliable downstream analysis. Non-rigid registration can calibrate warping tissue but cannot eliminate the alignment error.

Our method is robust as the convergence of CPD method is mainly rely on regional cell densities. We have demonstrated that our method is portable to different segmentation methods. Although the graph matching step works on strict conditions, it enables the possibility of introducing advanced technique such as graph neural network to measure the similarity between nodes and further establish cell correspondence. Since our alignment method relies on cell segmentations, the number of cells largely impact the effectiveness of applying our method to whole slide image (Supplementary Figure 4A), in which there could be hundreds of times more cells than a TMA core. Although CPD can be accelerated using specific techniques[9], it is still challenging to apply it to whole slide images due to the stress to the limits to the computation memory. To make our method applicable to whole slide image, we extended our method with Super-cell concept [22] by clustering cells based on proximity before applying our method (Supplementary Figure 4B,C). According to our preliminary evaluation, the alignment achieved comparable accuracy (Supplementary Figure D) compared to that with TMAs, which implies our method could be applied to whole slide images. However, we believe that histopathology image alignment is more meaningful for TMA slide scans. As compared to whole slide images, TMA slide scanning is a more realistic imaging method for a large cohort study [23].It only takes two landmarks to annotate each image pair to estimate the transformation defined in formular (1), so the annotation workload for whole slide image alignment is significantly lower than TMA core images. Moreover, regional distortion and tissue folding are more likely to happen in whole slide image. Even if we can regionally calibrate the alignment results, it brings extra challenges to seamlessly stitch the transformed images back into the new whole slide image.

Our cell feature concordance evaluation results suggested that H&E and MxIF morphological features are not always highly correlated. Based on our pathologists' observation, morphology features are more reliable in H&E. In our dataset, some cells detected in MxIF were fragments according to H&E. Meanwhile, some cells detected as one cell in MxIF but actually two/three cells in H&E, because some cells have strong fluorescence signals and spill over to adjacent cells. The discrepancies within cell segmentation introduced uncertainty to combined downstream analysis, but enabled another way to cross validate image qualities once the image pairs were aligned.

**Author contributions**

Jun Jiang, Markovic Svetomir and Chen Wang conceived of the presented idea. Jun Jiang and Raymond Moore developed the theory and performed the computations. Brenna Novotny, Leo Liu and Zachary Fogarty verified the analytical methods. Ray Guo, Chen Wang and Markovic Svetomir supervised the findings of this work. All authors discussed the results and contributed to the final manuscript.

**Data and Code availability**

Link to data: https://immunoatlas.org/MYCB/240802-1/MYCB24004/

Code and documentation can be found in this GitHub repository. https://github.com/dimi-lab/MultimodalityHistoComb

**Supplementary Figures:**

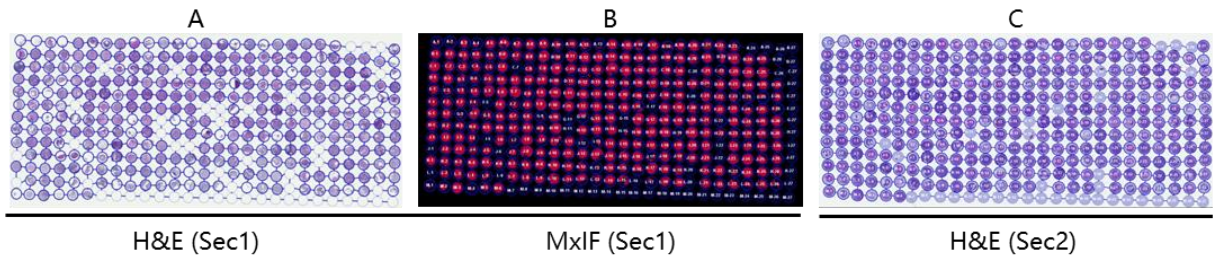

Supplementary Figure 1: Ovarian TMA dataset used in this study. A) and B) were a tissue section (Section 1) stained with MxIF protocol and then restained with H&E. C) was H&E stain of a serial section (Section 2) 5 µm apart from section 1.

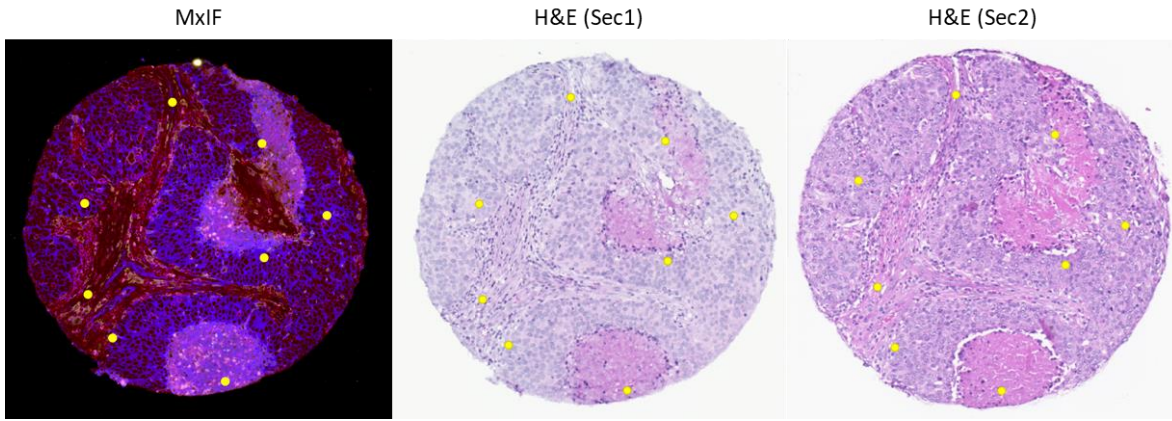

Supplementary Figure 2: TMA core landmark (eight points annotations). Each yellow dot was a landmark manually label for establishing ground truth transformation. The size of the landmarks was enlarged for better visualization. The actual resolution of landmarks equals to the pixel size of the image.

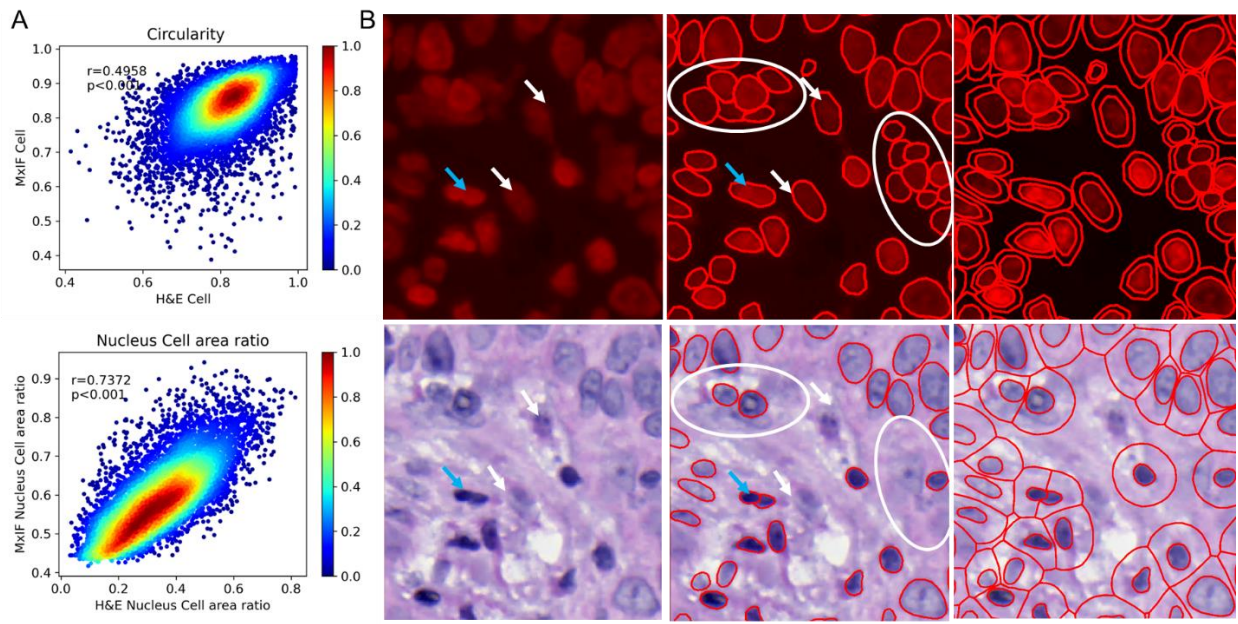

Supplementary Figure 3: Cell morphology differences caused by segmentation discrepancies. A) concordance of cell morphology features; B) segmentation differences within MxIF and H&E.

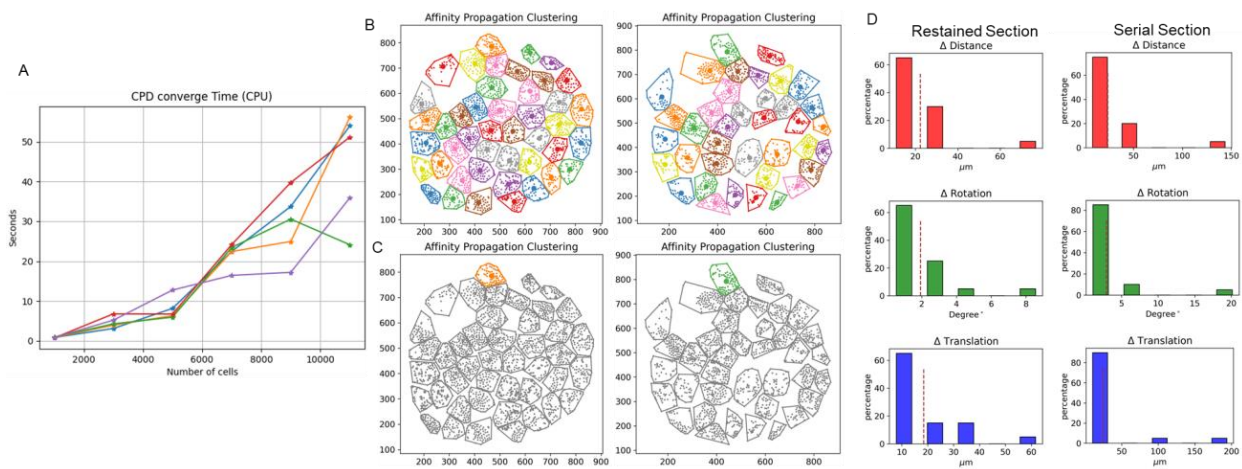

Supplementary Figure 4: A) Time efficiency of CPD. Each curve is an instance of simulation. B) Cell clustering results for a MxIF (left) and H&E (right) core. C) Clustering results with a pair of matched cell clusters. D) Quantitative evaluation results for restained and serial sections.